\documentclass[twoside,slac_one]{revtex4}
\usepackage{graphicx}
\usepackage{fancyhdr}
\usepackage{amsmath} 
\usepackage{bm}
\usepackage{amsxtra}
\usepackage{amssymb}
\usepackage{amsthm}
\usepackage{latexsym}
\usepackage{lscape}

\def\beq{\begin{equation}}
\def\eeq{\end{equation}}
\def\beqa{\begin{eqnarray}}
\def\eeqa{\end{eqnarray}}

\begin{document}

\title{Top Quark Theoretical Cross Sections and $p_T$ and Rapidity 
Distributions}

%

\author{Nikolaos Kidonakis}
\affiliation{Kennesaw State University, Physics \#1202, Kennesaw, GA 30144, USA}

\begin{abstract}
I present theoretical results for the top quark pair total cross section, 
and for the top quark transverse momentum and rapidity distributions 
at Tevatron and LHC energies. I also present results for single top quark 
production in the $t$- and $s$-channels and also via associated production 
with a $W$ boson. The calculations include approximate NNLO corrections that 
are derived from NNLL soft-gluon resummation.
\end{abstract}

\maketitle

\thispagestyle{fancy}


\section{Introduction}

\begin{figure}[h]
\centering
\includegraphics[width=35mm]{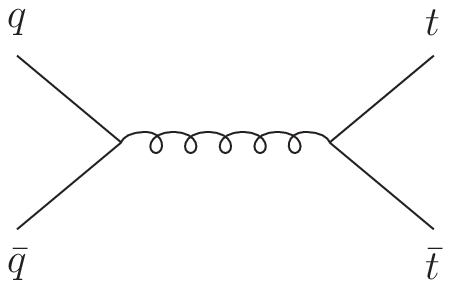}
\hspace{15mm}
\includegraphics[width=117mm]{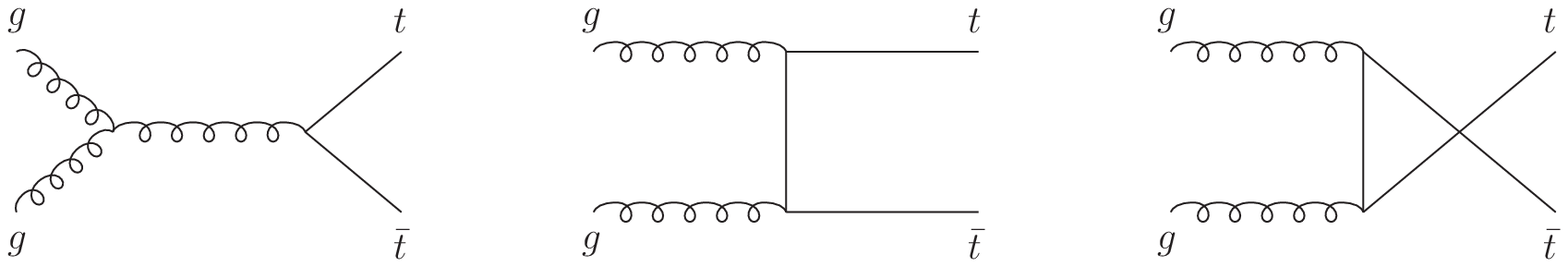}
\caption{LO $q{\bar q} \rightarrow t{\bar t}$ and $gg \rightarrow t{\bar t}$ 
diagrams.}
\label{qqgglo}
\end{figure}

The top quark is a very important part of the collider programs at the 
Tevatron and the LHC. Top quarks have been produced via top-antitop 
pair production and also via single top production at both colliders. 
The partonic processes at LO for top-antitop pair production are 
$q{\bar q} \rightarrow t {\bar t}$, dominant at the Tevatron, 
and $gg \rightarrow t {\bar t}$, dominant at the LHC.
The leading-order (LO) diagrams are shown in Fig. \ref{qqgglo}.

\begin{figure}[h]
\centering
\includegraphics[width=34mm]{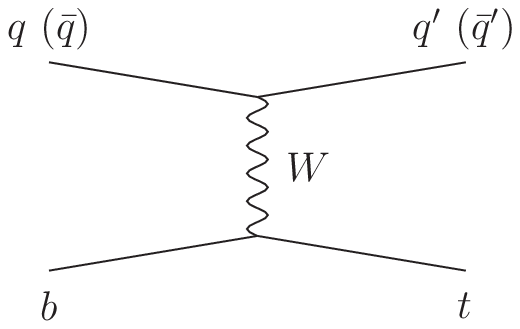}
\hspace{8mm}
\includegraphics[width=34mm]{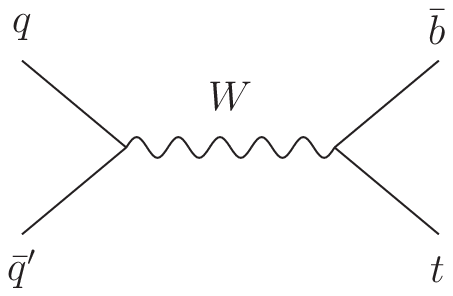}
\hspace{8mm}
\includegraphics[width=80mm]{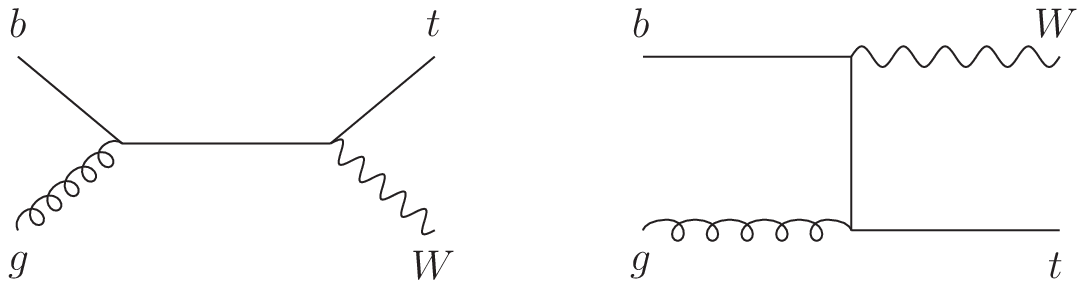}
\caption{LO $t$-channel, $s$-channel, and $tW$ diagrams.}
\label{tsWlo}
\end{figure}

Single top quark production can proceed via the $t$-channel processes, 
the $s$-channel processes, and via associated production of a top quark 
with a $W$ boson. The LO diagrams for these proceses are shown in 
Fig.~\ref{tsWlo}. 

The $t$-channel processes, $qb \rightarrow q' t$ and 
${\bar q} b \rightarrow {\bar q}' t$, 
are dominant at both the Tevatron and the LHC.
The $s$-channel processes,  $q{\bar q}' \rightarrow {\bar b} t$,  are
small at the Tevatron and LHC.
The associated production of a top quark with a $W$ boson, 
$bg \rightarrow tW^-$, has a very small cross section at the Tevatron, 
but is significant at the LHC.
A related process to $tW$ production is the associated production 
of a top quark with a charged Higgs boson in the 
Minimal Supersymmetric Standard Model, $bg \rightarrow tH^-$.

In Sec. 2 we discuss higher-order corrections for both $t{\bar t}$ and 
single-top production and identify soft-gluon corrections which are 
important near partonic threshold and contribute the dominant 
higher-order corrections in the hadronic cross section.
In Sec. 3 we give results for the total top-antitop pair cross section.
In Sec. 4 we show results for the top quark transverse momentum, $p_T$,  
distribution. In Sec. 5 we give results for the top quark rapidity, $Y$,  
distribution and the top quark forward-backward asymmetry. 
In Sec. 6 we present results 
for single top quark production via the $t$ channel, in Sec. 7 via the $s$ 
channel,  and in Sec. 8 via associated production with a $W$. We conclude 
in Sec. 9 with a summary.

\section{Higher-order corrections}

QCD corrections are significant for top pair and single top quark production
and are known fully at next-to-leading order (NLO).
Soft-gluon corrections from emission of soft (low-energy) gluons are an 
important contributor to the QCD corrections.
These soft-gluon corrections are dominant near threshold and they are 
of the form $[\ln^k(s_4/m^2)/s_4]_+$ 
with $k \le 2n-1$ and $s_4$ the kinematical distance from threshold.

We can resum these soft corrections via factorization and 
renormalization-group evolution.
At next-to-leading-logarithm (NLL) accuracy this requires one-loop  
calculations in the eikonal approximation.
Complete results now exist at next-to-next-to-leading-logarithm (NNLL) 
accuracy, using the two-loop soft anomalous dimension matrices for the 
corresponding partonic subprocesses.

An approximate next-to-next-to-leading-order (NNLO) cross section 
is derived from the expansion of the resummed cross section 
\cite{NKtop,NKdpf2011}.
The calculation is at the differential cross section level using 
single-particle-inclusive (1PI) kinematics. We note that 
1PI kinematics refer to partonic threshold, not just absolute threshold.

The threshold approximation works very well not only for Tevatron 
but also for LHC energies because partonic threshold is still important 
at the LHC:
there is only a 1\% difference between first-order approximate and exact 
corrections which translates into less than 1\% difference between NLO 
approximate and exact cross sections.

For our best prediction for the total cross section and differential 
distributions we add the NNLO approximate corrections from NNLL resummation
to the exact NLO results. 

\section{$t{\bar t}$ cross section}

\begin{figure}[h]
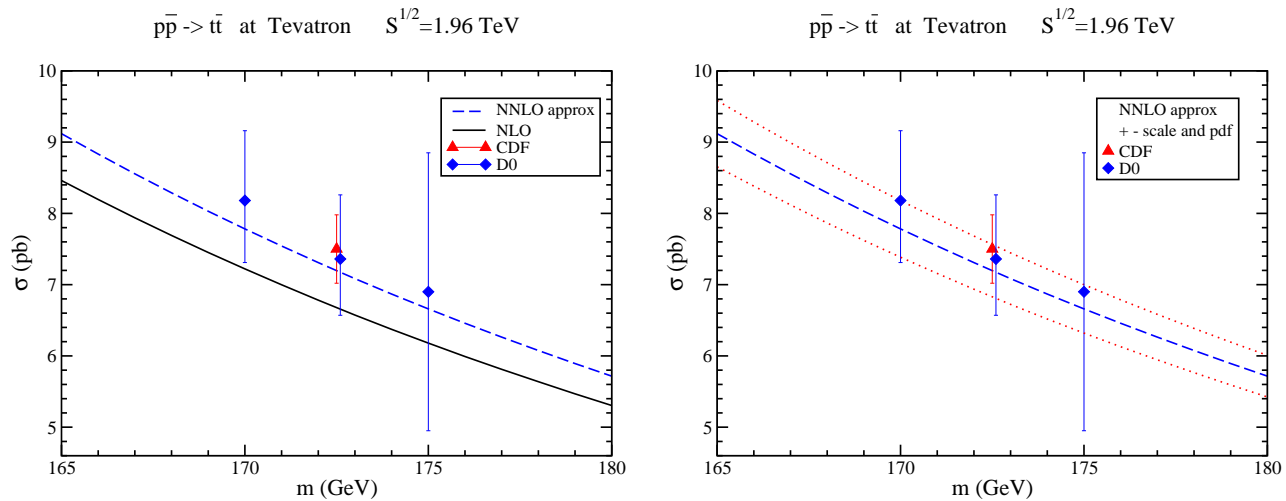

\centering
\includegraphics[width=82mm]{toptevexporder2011plot.eps}
\hspace{3mm}
\includegraphics[width=82mm]{toptevexp2011plot.eps}
\caption{Cross section for $t{\bar t}$ production at the Tevatron.}
\label{ttbarTevatron}
\end{figure}

We begin with the total top-antitop pair cross section. 
In Fig. \ref{ttbarTevatron} we show results for the exact NLO and the 
approximate NNLO cross section \cite{NKtop} at Tevatron energy and compare with 
recent results from the CDF \cite{CDFtt} and D0 \cite{D0tt} Collaborations.
The left plot shows that the NNLO approximate cross section 
describes the data better than NLO, while the plot on the right shows the 
approximate NNLO result together with the theoretical uncertainty, which is 
derived by varying the scale by a factor of two around the central value 
$\mu=m$ and adding this uncertainty in quadrature with the uncertainty 
from the MSTW2008 NNLO parton distribution functions (pdf) \cite{MSTW}  
at 90\% C.L.

The $t{\bar t}$ cross section at the Tevatron for a top quark mass of 173 GeV
is 
\beqa
\hspace{1cm}\sigma^{\rm NNLOapprox}_{t{\bar t}}(m_t=173 \, {\rm GeV}, \, 1.96\, {\rm TeV})&=&7.08 {}^{+0.00}_{-0.24} {}^{+0.36}_{-0.27} \; {\rm pb}
\nonumber
\eeqa
where the first uncertainty is from scale variation and the second is from 
the pdf.
The NNLO approximate corrections enhance the cross section by 7.8\% 
over the NLO result (with the same pdf).

\begin{figure}[h]
\centering
\includegraphics[width=100mm]{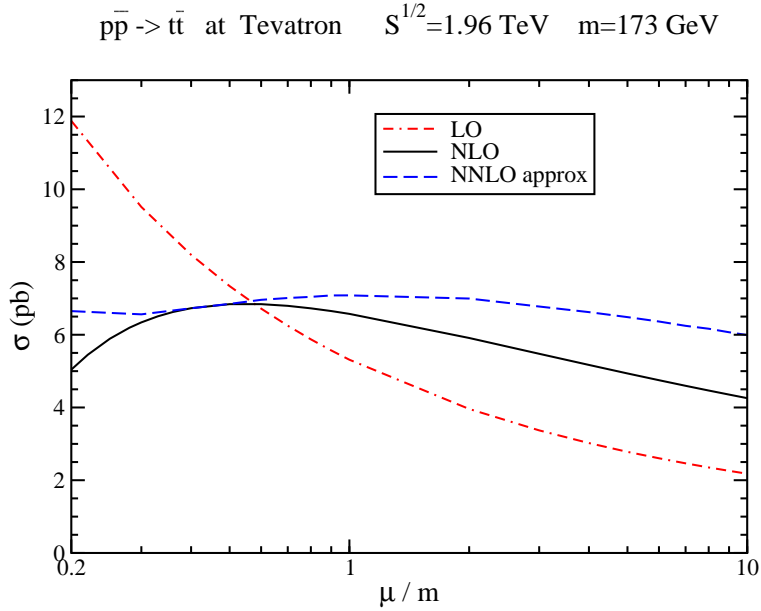}
\caption{Scale dependence of the $t{\bar t}$ cross section at the Tevatron.}
\label{ttbarmuTev}
\end{figure}

In Fig. \ref{ttbarmuTev} we show the scale dependence of the $t{\bar t}$ 
cross section at the Tevatron. The plot shows that the LO result 
displays a large dependence on the scale. The NLO cross section 
has a much milder dependence 
and the NNLO approximate result significantly reduces the dependence even 
further. 

\begin{figure}[h]
\centering
\includegraphics[width=100mm]{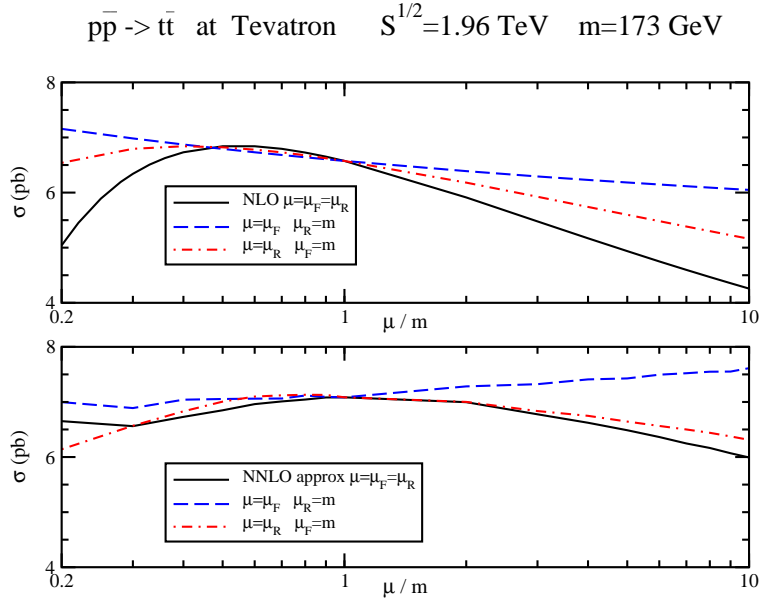}
\caption{NLO and NNLO $\mu_F$ and $\mu_R$ dependence of the $t{\bar t}$
cross section at the Tevatron.}
\label{ttbarmuFRTev}
\end{figure}

Figure \ref{ttbarmuFRTev} shows the separate dependence of the NLO 
and approximate NNLO cross sections on independent variations of the 
factorization scale $\mu_F$ and the renormalization scale $\mu_R$.
We observe a reduction at NNLO in the independent scale variations as well.

\begin{figure}[h]
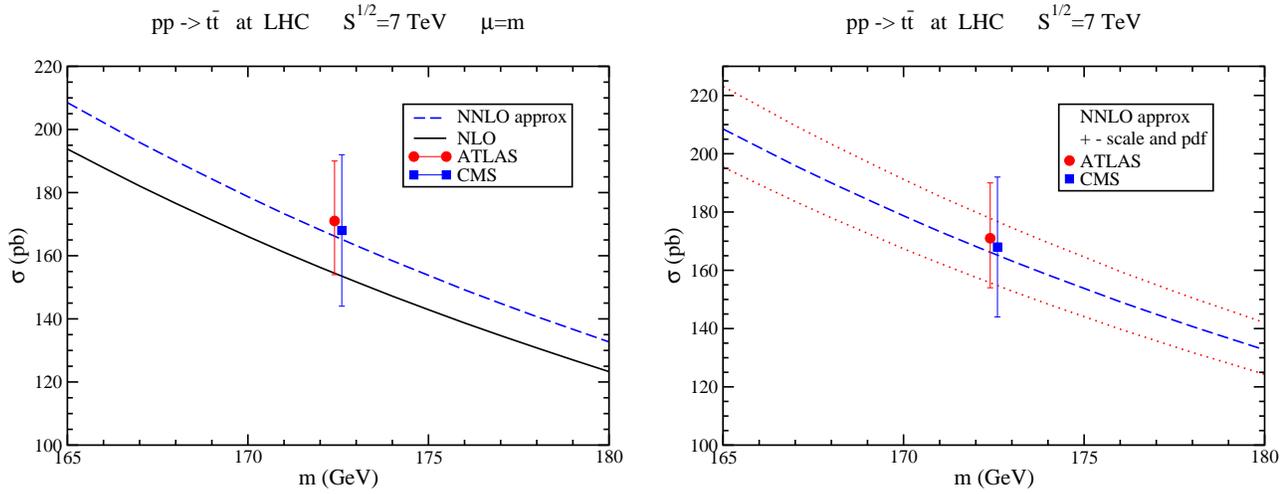

\centering
\includegraphics[width=82mm]{top7lhcexporder2011plot.eps}
\hspace{3mm}
\includegraphics[width=82mm]{top7lhcexp2011plot.eps}
\caption{Cross section for $t{\bar t}$ production at the LHC.}
\label{ttbarLHC}
\end{figure}

In Fig. \ref{ttbarLHC} we show the $t{\bar t}$ cross section at the LHC 
at 7 TeV energy and compare with recent results from the ATLAS \cite{ATLAStt} 
and CMS \cite{CMStt} Collaborations. The left plot shows that the NNLO 
approximate cross section is in better agreement with the data 
than the NLO result. 
The plot on the right shows the central NNLO approximate cross section 
as before together with the uncertainty from scale variation and pdf errors.

The $t {\bar t}$ cross section for a top quark mass of 173 GeV at the LHC 
with 7 TeV energy is 
\beqa
\hspace{1cm} \sigma^{\rm NNLOapprox}_{t{\bar t}}(m_t=173\, {\rm GeV}, \, 7\, {\rm TeV})&=&163 {}^{+7}_{-5}  {}^{+9}_{-9} \; {\rm pb}
\nonumber 
\eeqa
while the corresponding result at 14 TeV energy is
\beqa
\hspace{1.1cm} \sigma^{\rm NNLOapprox}_{t{\bar t}}(m_t=173\, {\rm GeV}, 14\, {\rm TeV})&=&920 {}^{+50}_{-39}{}^{+33}_{-35} \; {\rm pb} \, .
\nonumber 
\eeqa

The NNLO approximate corrections contribute an enhancement over NLO of 7.6\% 
at 7 TeV and 8.0\% at 14 TeV. 

\begin{figure}[h]
\centering
\includegraphics[width=110mm]{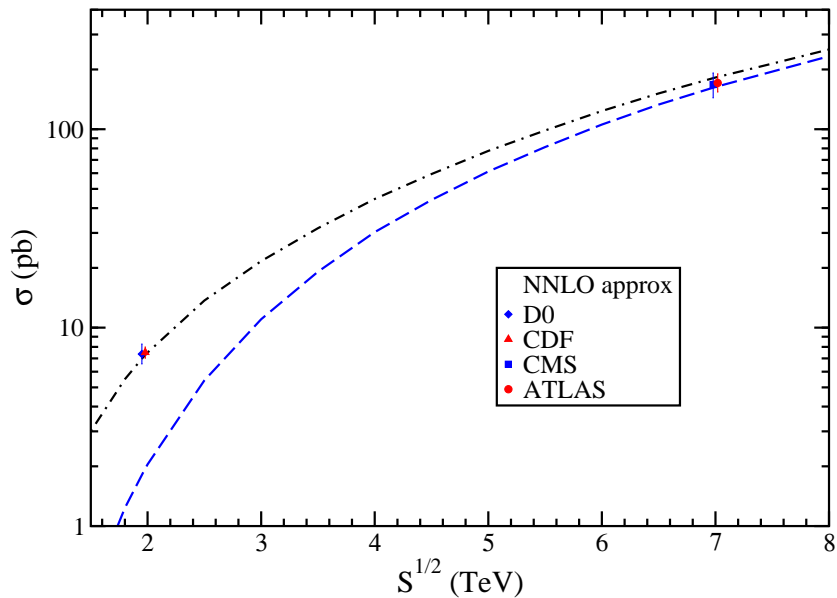}
\caption{Cross section for $t{\bar t}$ production at 
$p{\bar p}$ and $pp$ colliders.}
\label{ttbarS}
\end{figure}

Figure \ref{ttbarS} plots the $t{\bar t}$ theoretical cross sections versus 
collider energy for $p{\bar p}$ and $pp$ collisions 
together with experimental measurements of the cross section. 
The Tevatron and LHC data are in very good agreement with theory.

\section{Top quark transverse momentum distributions}

\begin{figure}[h]
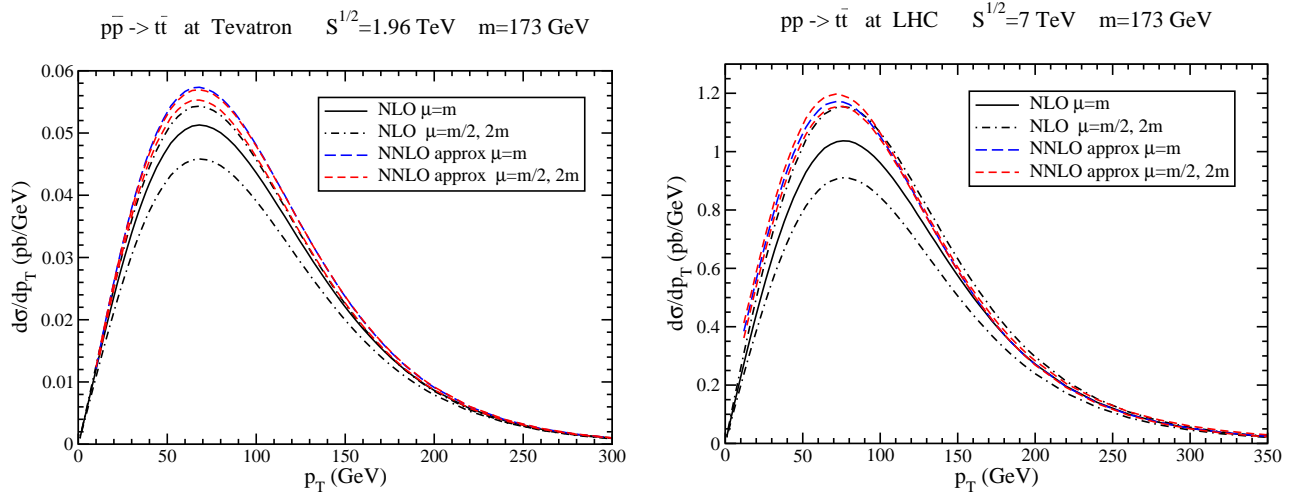

\centering
\includegraphics[width=82mm]{pttev1plot.eps}
\hspace{3mm}
\includegraphics[width=82mm]{pt7lhc1plot.eps}
\caption{Top quark $p_T$ distribution at the Tevatron and the LHC.}
\label{toppT}
\end{figure}

The top quark transverse momentum distribution, $d\sigma/dp_T$, 
at the Tevatron and at the LHC at 7 TeV is plotted in Fig. \ref{toppT}. 
NLO and approximate NNLO results \cite{NKtop}
are displayed for the central scale value $\mu=m$ as well as for 
$\mu=m/2$ and $2m$. 

The NNLO soft-gluon corrections enhance the $p_T$
distribution but they do not significantly change the shape in the $p_T$ range
shown. The scale dependence is reduced relative to NLO, in line with the
reduction that was observed for the total cross section.

\section{Top quark rapidity distributions and forward-backward asymmetry}

\begin{figure}[h]
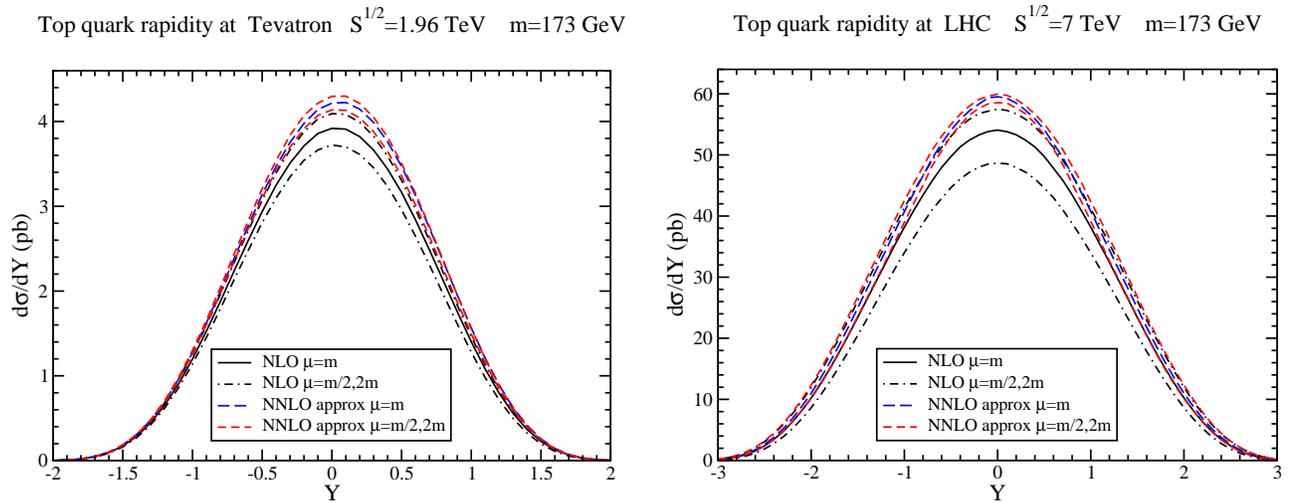

\centering
\includegraphics[width=82mm]{ytevplot.eps}
\hspace{3mm}
\includegraphics[width=82mm]{y7lhcplot.eps}
\caption{Top quark rapidity distribution at the Tevatron and the LHC.}
\label{topY}
\end{figure}

The top quark rapidity distribution, $d\sigma/dY$, at the Tevatron and at the LHC
at 7 TeV is plotted in Fig. \ref{topY}. Again, NLO and approximate NNLO results 
\cite{NKrap} are displayed for the central scale value $\mu=m$ as well as for 
$\mu=m/2$ and $2m$. The NNLO soft-gluon corrections enhance the rapidity
distribution and reduce the scale dependence relative to NLO but they do not 
significantly change the shape in the rapidity range shown. 

The rapidity distribution enters into the definition of the 
forward-backward asymmetry,
$$
A_{\rm FB}=\frac{\sigma(Y>0)-\sigma(Y<0)}{\sigma(Y>0)+\sigma(Y<0)} \, .
$$

The $gg$ channel is symmetric at all orders.
The $q{\bar q}$ channel on the other hand is asymmetric starting at NLO. 
The asymmetry is significant at the Tevatron. The 
theoretical result for the top quark asymmetry at the Tevatron at approximate NNLO 
is $A_{\rm FB}=0.052^{+0.000}_{-0.006}$ where the uncertainty indicated 
is from scale variation. The theoretical value is much smaller than current 
experimental values.

\section{Single top quark production - $t$ channel}

The $t$-channel production of a single top quark proceeds via the exchange 
of a spacelike $W$ boson. The process is dominant among single top channels 
at both Tevatron and LHC energies. 

For a top quark mass of 173 GeV the approximate NNLO $t$-channel single top cross sections \cite{NKtch} are
\beqa
\sigma^{\rm NNLOapprox,\, top}_{t-{\rm channel}}(m_t=173 \, {\rm GeV}, \, 1.96\, {\rm TeV})&=&1.04 {}^{+0.00}_{-0.02} \pm 0.06 \; {\rm pb}
\nonumber \\
\sigma^{\rm NNLOapprox,\, top}_{t-{\rm channel}}(m_t=173 \, {\rm GeV}, \, 7\, {\rm TeV})&=&41.7 {}^{+1.6}_{-0.2} \pm 0.8 \; {\rm pb}
\nonumber \\
\sigma^{\rm NNLOapprox,\, top}_{t-{\rm channel}}(m_t=173 \, {\rm GeV}, \, 14\, {\rm TeV})&=&151 {}^{+4}_{-1} \pm 3 \; {\rm pb} \, .
\nonumber
\eeqa

For $t$-channel single antitop production the cross section at the Tevatron is 
the same as for top.
However, the single antitop production cross section at the LHC in the  $t$ 
channel is different:
\beqa
\sigma^{\rm NNLOapprox,\, antitop}_{t-{\rm channel}}(m_t=173 \, {\rm GeV}, \, 7\, {\rm TeV})&=&22.5 \pm 0.5 {}^{+0.7}_{-0.9} \; {\rm pb}
\nonumber \\
\sigma^{\rm NNLOapprox,\, antitop}_{t-{\rm channel}}(m_t=173 \, {\rm GeV}, \, 14\, {\rm TeV})&=&92 {}^{+2}_{-1} {}^{+2}_{-3} \; {\rm pb} \, .
\nonumber
\eeqa

\begin{figure}[h]
\centering
\includegraphics[width=110mm]{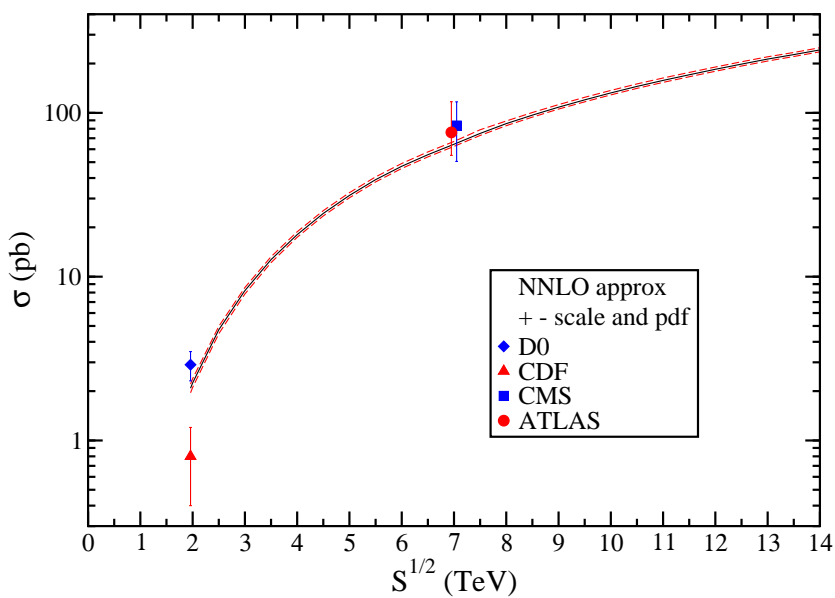}
\caption{$t$-channel cross section at hadron colliders.}
\label{tchtop}
\end{figure}

The $t$-channel combined cross section (single top + single antitop) 
is plotted versus energy in Fig. \ref{tchtop} together with recent measurements
at the Tevatron \cite{CDFt,D0t} and the  LHC \cite{ATLASt,CMSt}. 
Again the Tevatron and LHC results are consistent with theory.

\section{Single top quark production - $s$ channel}

The $s$-channel production of a single top quark proceeds via the exchange 
of a timelike $W$ boson. This process has the smallest cross section among 
single top channels at LHC energies. 

For a top quark mass of 173 GeV the approximate NNLO $s$-channel single top 
cross sections \cite{NKsch} are
\beqa
\sigma^{\rm NNLOapprox,\, top}_{s-{\rm channel}}(m_t=173 \, {\rm GeV}, \, 1.96\, {\rm TeV})&=&0.523 {}^{+0.001}_{-0.005} {}^{+0.030}_{-0.028} \; {\rm pb}
\nonumber \\
\sigma^{\rm NNLOapprox,\, top}_{s-{\rm channel}}(m_t=173\, {\rm GeV}, \, 7\, {\rm TeV})&=&3.17 \pm 0.06 {}^{+0.13}_{-0.10} \; {\rm pb}
\nonumber \\
\sigma^{\rm NNLOapprox,\, top}_{s-{\rm channel}}(m_t=173\, {\rm GeV}, 14\, {\rm TeV})&=&7.93 \pm 0.14 {}^{+0.31}_{-0.28} \; {\rm pb} \, .
\nonumber 
\eeqa
The NNLO approximate corrections provide an enhancement over NLO 
(with the same pdf) of 15\% at the Tevatron and 13\% at the LHC, and thus 
they have a very significant impact. 

The $s$-channel single antitop production cross section at the Tevatron is 
the same as for top.
However, the $s$-channel single antitop production cross section 
at the LHC is different:
\beqa
\sigma^{\rm NNLOapprox,\, antitop}_{s-{\rm channel}}(m_t=173\, {\rm GeV}, \, 7\, {\rm TeV})&=&1.42 \pm 0.01 {}^{+0.06}_{-0.07} \; {\rm pb}
\nonumber \\
\sigma^{\rm NNLOapprox,\, antitop}_{s-{\rm channel}}(m_t=173\, {\rm GeV}, 14\, {\rm TeV})&=&3.99 \pm 0.05 {}^{+0.14}_{-0.21} \; {\rm pb} \, .
\nonumber 
\eeqa

\section{Associated $tW^-$ production}

The associated production of a top quark with a $W$ boson,
$bg \rightarrow tW^-$, is negligible at the Tevatron but important 
at the LHC.
For a top quark mass of 173 GeV the approximate NNLO cross section \cite{NKtW} 
is 
\beqa
\sigma^{\rm NNLOapprox}_{tW}(m_t=173 \, {\rm GeV}, \, 7\, {\rm TeV})&=&7.8 \pm 0.2 {}^{+0.5}_{-0.6} \; {\rm pb}
\nonumber \\
\sigma^{\rm NNLOapprox}_{tW}(m_t=173 \, {\rm GeV}, 14\, {\rm TeV})&=&41.8 \pm 1.0 {}^{+1.5}_{-2.4} \; {\rm pb} \, .
\nonumber 
\eeqa
The NNLO approx corrections increase the NLO cross section by $\sim 8$\%.
The cross section for ${\bar t}W^+$ production is identical to that for top.

A related process is the associated production of a top quark with a charged 
Higgs in the Minimal Supersymmetric Standard Model. In that case 
the NNLO approximate corrections increase the NLO cross section by 
$\sim 15$ to $\sim 20$\%. Another related process is $W$ production at 
large $p_T$ \cite{NKdpf2011,KGS}. 

\section{Summary}

We have discussed NNLL soft-gluon resummation for top quark pair and 
single top 
production. We have derived NNLO approximate cross sections from the 
expansion of the NNLL resummed cross section.  
Numerical results for the $t {\bar t}$ production cross section 
have been presented at both Tevatron and LHC energies. The NNLO 
approximate corrections reduce the scale dependence of the cross section.

The top quark $p_T$ and rapidity distributions have also been presented 
and, again, the NNLO soft-gluon corrections enhance the cross section 
and reduce the scale dependence but they do not significantly affect 
the shape of the distributions.
The theoretical top quark forward-backward asymmetry has also been 
calculated and is found to be significantly smaller 
than observed at the Tevatron.

The cross sections for $t$-channel and $s$-channel single top production 
as well as the associated production of a top quark with a 
$W$ boson have also been presented at both Tevatron and LHC.

The NNLO approximate corrections for top pair and single top 
production are significant at both the Tevatron and LHC colliders 
and they reduce the theoretical uncertainty. The theoretical results 
are in good agreement with experimental measurements from CDF and 
D0 at the Tevatron, and from ATLAS and CMS at the LHC. 

\begin{acknowledgments}
This work was supported by the National Science Foundation under 
Grant No. PHY 0855421.
\end{acknowledgments}

\bigskip 

\end{document}